\documentclass[conference]{IEEEtran}
\IEEEoverridecommandlockouts

\usepackage{cite}
\usepackage{amsmath,amssymb,amsfonts}
\usepackage{algorithmic}
\usepackage{graphicx}
\usepackage{textcomp}
\usepackage{booktabs}
\usepackage[dvipsnames]{xcolor}
\usepackage{soul}
\usepackage{url}
\usepackage[dvipsnames]{xcolor}
\def\BibTeX{{\rm B\kern-.05em{\sc i\kern-.025em b}\kern-.08em
    T\kern-.1667em\lower.7ex\hbox{E}\kern-.125emX}}
\makeatletter
\newcommand{\linebreakand}{%
  \end{@IEEEauthorhalign}
  \hfill\mbox{}\par
  \mbox{}\hfill\begin{@IEEEauthorhalign}
}
\makeatother

\begin{document}

% ------------------------- TITLE ----------------------
\title{Enabling Time-Aware Priority Traffic Management over Distributed FPGA Nodes
%{\large \textbf{(Short Paper)}}
}

% ------------------- AUTHOR LIST ----------------------
% LINKS Foundation
% Author list -- basic
\author{\IEEEauthorblockN{Alberto Scionti}
\IEEEauthorblockA{\textit{CPE Research Group} \\ \textit{LINKS Foundation}, ITALY \\
alberto.scionti@linksfoundation.com}
\and
\IEEEauthorblockN{Paolo Savio}
\IEEEauthorblockA{\textit{CPE Research Group} \\ \textit{LINKS Foundation}, ITALY \\
paolo.savio@linksfoundation.com}
\and
\IEEEauthorblockN{Francesco Lubrano}
\IEEEauthorblockA{\textit{CPE Research Group} \\ \textit{LINKS Foundation}, ITALY \\
francesco.lubrano@linksfoundation.com}
\linebreakand
\IEEEauthorblockN{Federico Stirano}
\IEEEauthorblockA{\textit{CPE Research Group} \\ \textit{LINKS Foundation}, ITALY \\
federico.stirano@linksfoundation.com}
\and
\IEEEauthorblockN{Antonino Nespola}
\IEEEauthorblockA{\textit{CPE Research Group} \\ \textit{LINKS Foundation}, ITALY \\
antonino.nespola@linksfoundation.com}
\and
\IEEEauthorblockN{Olivier Terzo}
\IEEEauthorblockA{\textit{CPE Research Group} \\ \textit{LINKS Foundation}, ITALY \\
olivier.terzo@linksfoundation.com}
\linebreakand
% Politecnico di Torino
\IEEEauthorblockN{Corrado De Sio}
\IEEEauthorblockA{\textit{Dept. Control and Computer Engineering} \\ \textit{Politecnico di Torino}, ITALY \\
corrado.desio@polito.it}
\and
\IEEEauthorblockN{Luca Sterpone}
\IEEEauthorblockA{\textit{Dept. Control and Computer Engineering} \\ \textit{Politecnico di Torino}, ITALY \\
luca.sterpone@polito.it}
}

\maketitle

% ---------------------- ABSTRACT ----------------------
% Length is max. 300 words 
% ------------------------------------------------------
\begin{abstract}
Network Interface Cards (NICs) greatly evolved from simple basic devices moving traffic in and out of the network to complex heterogeneous systems offloading host CPUs from performing complex tasks on in-transit packets. These latter comprise different types of devices, ranging from NICs accelerating fixed specific functions (e.g., on-the-fly data compression/decompression, checksum computation, data encryption, etc.) to complex Systems-on-Chip (SoC) equipped with both general purpose processors and specialized engines (Smart-NICs). Similarly, Field Programmable Gate Arrays (FPGAs) moved from pure reprogrammable devices to modern heterogeneous systems comprising general-purpose processors, real-time cores and even AI-oriented engines. Furthermore, the availability of high-speed network interfaces (e.g., SFPs) makes modern FPGAs a good choice for implementing Smart-NICs. In this work, we extended the functionalities offered by an open-source NIC implementation (Corundum) by enabling time-aware traffic management in hardware, and using this feature to control the bandwidth associated with different traffic classes. By exposing dedicated control registers on the AXI bus, the driver of the NIC can easily configure the transmission bandwidth of different prioritized queues. Basically, each control register is associated with a specific transmission queue (Corundum can expose up to thousands of transmission and receiving queues), and sets up the fraction of time in a transmission window which the queue is supposed to get access the output port and transmit the packets. Queues are then prioritized and associated to different traffic classes through the Linux QDISC mechanism. Experimental evaluation demonstrates that the approach allows to properly manage the bandwidth reserved to the different transmission flows.
\end{abstract}

% ---------------------- KEYWORDS ----------------------

\section{Introduction}
\label{sec:introduction}
The interconnection subsystem is of primary importance to enable end users extracting the best performance from a distributed computer system, and many factors contribute to define its characteristics (e.g., topology, supported protocols, etc.).
%Many factors contribute to define the interconnection characteristics of a distributed system, like the radix of each node in the communication network, the arrangement of nodes in the overall system (topology), the supported protocols and the communication speed. 
In a distributed system, network interface cards (NICs) represent one of the main building blocks of the interconnection, being the primary responsible for managing ingress and egress traffic from a compute node. NICs evolved over the years, from simple devices responsible for moving in and out packets from the local compute node (host system), to more advanced devices offloading the host CPU from performing most of the packet processing tasks. Recently, NICs further evolved to embed local processing capabilities, through dedicated acceleration engines and general purpose cores, leading to what is generally referable to as \textit{Smart-NICs} and as \textit{data processing units} (DPUs).

Field Programmable Gate Arrays (FPGAs) represent a valuable tool for experimenting with hardware and software co-design and for fast-prototyping complex digital systems, as well as testing new hardware features without incurring into large costs of manufacturing complex application specific ICs (ASICs). Modern FPGAs are effective heterogeneous systems integrating large programmable logic resources (Look-Up Tables, Flip-Flops, local RAM blocks, etc.), energy-efficient general purpose cores, and AI-tailored and real-time cores (System-on-Chip FPGAs). Moreover, they offer high-speed communication interfaces, with line rates up to 100~Gbps per port. As such, FPGAs are a good target for implementing Smart-NICs over a (large-scale) distributed computer system, where packet processing and related data processing tasks can be effectively executed on the same devices. Example of large deployment of FPGAs can be found in Cloud providers~\cite{b01, b02, b03}, and supercomputers~\cite{b04, b15, b27}.

%There are smart-NICs offered from almost all the vendors; in the majority of the cases, these products couple high-speed connectivity per port (up to 400~Gbps) with a large compute capabilities given by manycore and dedicated accelerator engines. To further extend their flexibility and programmability, these devices generally run full-fledged operating systems (e.g., Linux) along with dedicated programming frameworks. Despite this large versatility, smart-NICs available on the market does not provide enough flexibility and openness to allow for customization, as required by researchers. To overcome such limitation in the way hardware-software co-design can be explored for improving smart-NICs, open-source projects emerged. These latter, explore more the features offered by modern SoC FPGA to implement a fully customizable hardware-software solutions, which enable end-users to experimenting with the design of new protocols, network architectures and distributed applications.

In this work, we propose a mechanism to enable the management of traffic prioritization by enabling a time-aware scheduling of egress traffic, on top of a Smart-NIC architecture supporting multiple receiving (RX) and transmitting (TX) packet queues. We leverage on Corundum~\cite{b05}, an open-source project which can handle up to thousands of RX-TX packet queues, and which can be easily deployed on SoC FPGAs. We enabled time-aware hardware management of the queues, exporting configuration registers that allow the Linux kernel stack to precisely set the amount of transmission time for each managed queue. By configuring Linux QDISC in such a way the highest traffic priority is associated to traffic classes mapped on those TX queues that received larger time slots, we can control the bandwidth associated to each traffic priority. Experimental evaluation, on a distributed system composed of a group of SoC FPGA based nodes, confirms that with this approach we can let the users to easily control the outgoing traffic, not only in terms of priorities, but also in terms of allocated bandwidth.

The paper is organized as follows: Section~\ref{sec:introduction} provides some introductory material, Section~\ref{sec:backgrounds} provides foundational concepts concerning a generic NIC architecture; Section~\ref{sec:system_overview} describes our target system along with details on the addressing scheme and the software interface enhancing programmability. It follows Section~\ref{sec:implementation}, where we go in depth with our implementation and preliminary experimental evaluation. Section~\ref{sec:sota} describes State-of-the-Art, while Section~\ref{sec:conclusion} summarizes the work and provides some future research lines.

% sect. 2
\section{Background}
\label{sec:backgrounds}
\begin{figure}[ht]
  \centering
  \includegraphics[width=0.85\linewidth]{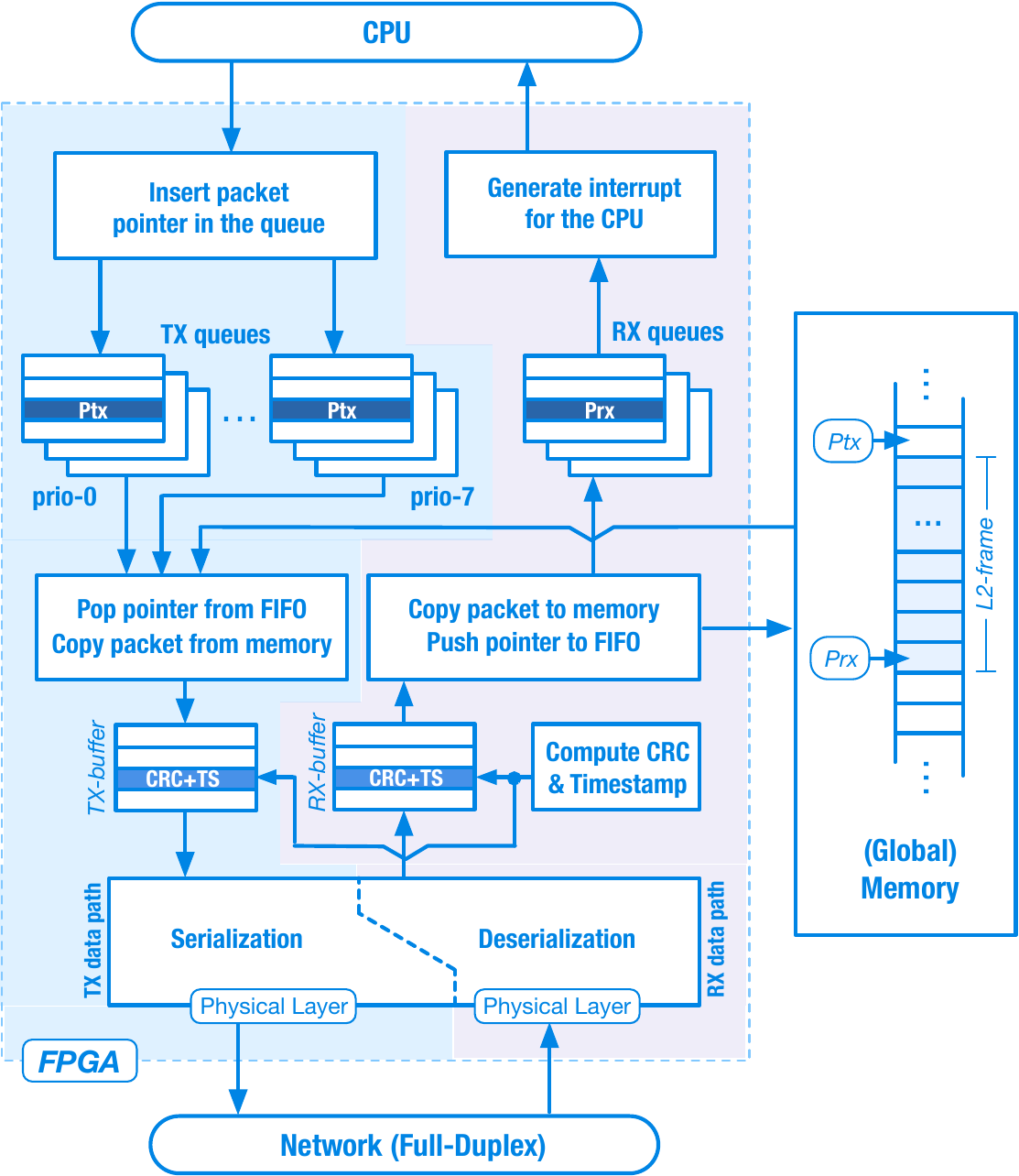}
  \caption{Architectural overview of the main components of a NIC; receiving (right side) and transmitting (left side) data paths are separated.}
    \label{fig:background_nic}
\end{figure}
% Questa sezione andrà rivista:
%  1. probabilmente, quasi tutto il contenuto che è indicato qui lo si potrà spostare sulla sezione State-of-the-Art
%  2. si può dedicare questa sezione a descrivere un po' come è fatta e funziona una NIC generica (e quindi come è fatto Corundum)

Despite network interface cards evolved over the years from simple devices to complex accelerator-like systems, their internal organization did not change very much. As such, the number of internal components increased to accommodate more data structures and increase performance, although the core functionalities remained almost unchanged. Figure~\ref{fig:background_nic} depicts the internal architecture of a generic NIC, highlighting the main core components that are part of the receiving (RX) and transmitting (TX) data paths. The physical layer enables a full-duplex communication and is responsible for adapting the transmission of the bitstream composing a L2 frame (e.g., an Ethernet frame) to the physical signalling of the network interface. For instance, using a small form-factor pluggable (SFP) interface along with an optical cable, the physical layer controls the signals used to drive the laser. Built on top of the physical layer, there is a hardware block that is primarily responsible for \textit{deserialising} (RX data path) the received bitstream or \textit{serialising} (TX data path) the content of a L2 frame in order to transmit it to the network. On the receiving side, a new L2 frame is written in a local buffer; in the meantime a dedicated block computes the CRC code and verifies the correctness of the received frame. In case the received CRC is different from the one computed, the frame is dropped; otherwise, a timestamp is added to the frame. On the transmitting side, this latter block can be used to compute the CRC of the frame and marking the timestamp at which the frame is actually transmitted.

Interestingly, modern NICs keep multiple internal FIFOs to improve the overall performance. Indeed, by having separated queues, the NIC can handle multiple concurrent packet flows independently, thus reducing contention, minimizing head-of-line blocking, and enabling parallelism in packet processing. These FIFOs are generally managed through a round-robin policy. Another approach introduced to increase the performance is that of copy-once the frame content (via DMA), while moving, within the system, the pointer (i.e., few bytes) to the memory location where the frame content has been copied. This implies that the main memory (there is a main memory region of the host system that is shared with the NIC) has a region organized as a circular buffer (these correspond to the circular buffers also depicted in figure~\ref{fig:time_aware_schedule}) and shared with the host CPU. For instance, in figure~\ref{fig:background_nic}, pointer $P_{rx}$ on the receiving side (similar situation is for pointers $P_{tx}$ in the transmitting side) is used to access the frame content. A dedicated block is responsible for copying the frame on the circular buffer and pushing the pointer in the selected FIFO. Asynchronously, the NIC hardware pops pointers from the head of the RX queues and generates interrupts for the host CPU. In the Linux environment, the interrupt management is optimized. As such, instead of taking care of an interrupt for each received frame, the kernel tries to get advantage from multicore architectures, by asking the driver to pop a group of frames and assigning them to multiple cores.

\begin{figure*}[ht]
  \centering
  \includegraphics[width=0.85\linewidth]{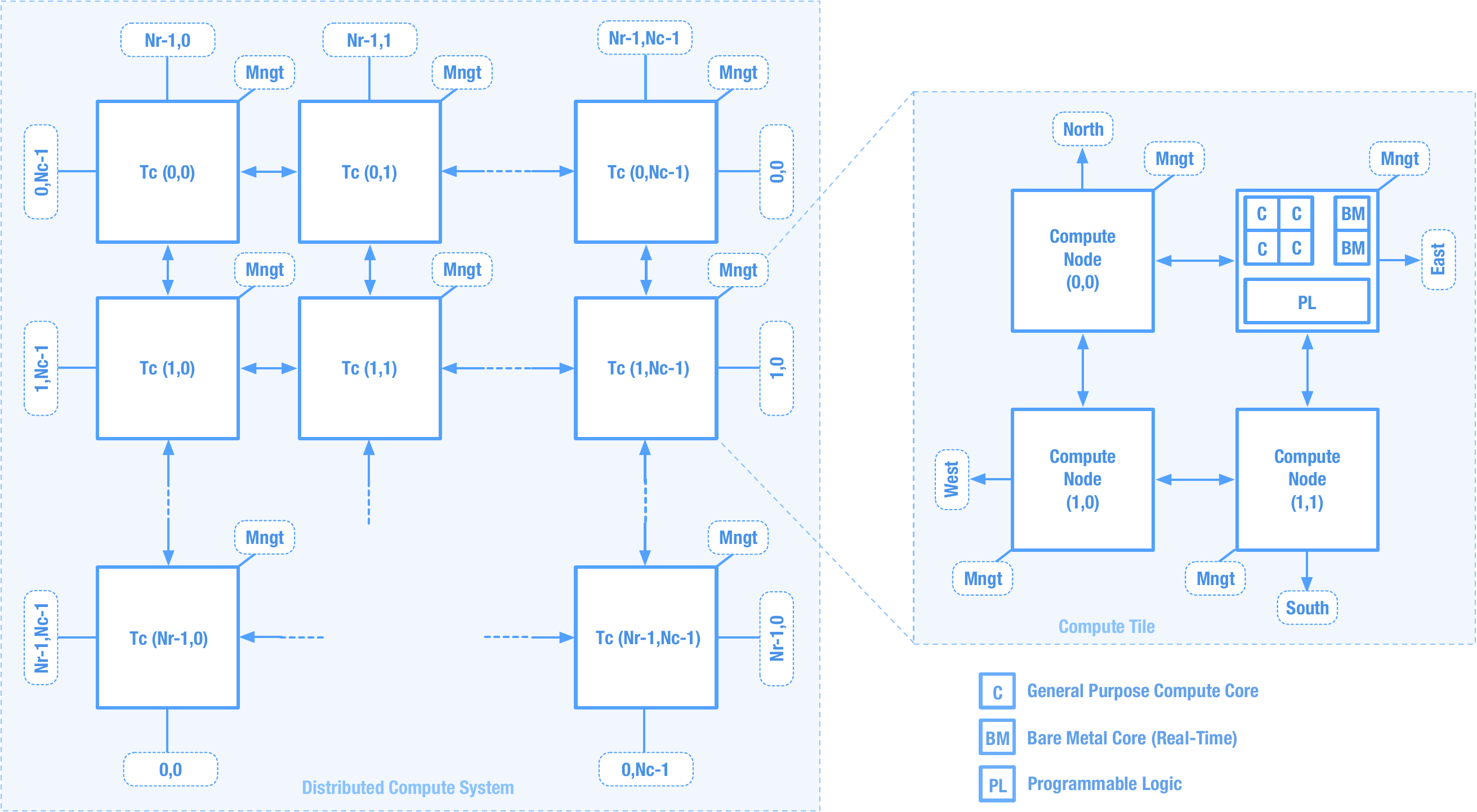}
  \caption{General overview of the target distributed system using SoC FPGAs as compute nodes.}
    \label{fig:system_overview}
\end{figure*}

On the transmitting side, the number of FIFOs can be large than that of the receiving side, since priorities must be managed. Given the frame priority, the selection of a FIFO is done in a round-robin fashion. A specific block is responsible to push in the FIFO the pointer to the frame content. Asynchronously, pointers are popped from the head of the FIFOs depending on their priorities. Here, the frame content is copied back into an internal transmission buffer. Similarly to what happens at the receiving side, CRC (here, the kernel can take advantage from the availability of dedicated hardware to offload this task) and timestamp are added to the frame. Serialization of the buffer content allows to stream out the frame to the network.

Finally, it is worth to note that in this architecture, the serialisation/deserialisation block along with that computing the CRC and adding the timestamps can replicated for each network port exposed by the NIC, while the management of FIFOs and the interface with the host CPU are shared resources.

% sect. 3
\section{System overview}
\label{sec:system_overview}
%
%\begin{figure*}[ht]
%  \centering
%  \includegraphics[width=0.80\linewidth]{figs/system_overview.pdf}
%  \caption{General overview of the target distributed system using SoC FPGAs as compute nodes.}
%    \label{fig:system_overview}
%\end{figure*}

Our target is a distributed system composed of groups of SoC FPGA based nodes, i.e., compute tiles $Tc$, which are organized as a hierarchical 2D-mesh. The mesh is extended to form a 2D-torus topology, which guarantees better performance over the simpler mesh. Each compute tile is composed by a subgroup of nodes, forming a simple 2D-mesh. Interestingly, although the mesh can be easily extended in both the horizontal and vertical dimensions, the minimum size of a tile is $2 \times 2$, since we exploit one port on each compute node to provide external connectivity for the $Tc$ in the horizontal dimension (east, west) and in the vertical dimension (north, south). Each compute node is formed by a multiprocessor subsystem containing both general-purpose and real-time (bare-metal) cores, and coupled with a programmable logic implementing the NIC. Each NIC exposes $Np$ network ports; since one port is dedicated to the external connectivity towards other compute tiles, $Np - 1$ ports remain available for the internal connectivity. However, we allocate one additional port for connecting the compute node to a separated management network; so, with a typical connectivity of 4 ports, our strategy led to limit the size of the compute tile to $|Tc| = 2 \times 2$ and to reserve two ports for connecting each compute node to each neighbourhoods. Figure~\ref{fig:system_overview} depicts the architecture of our target system.

Within this system, each \textit{compute node} exposes a NIC implemented on the programmable logic ($PL$) of the device, leaving the general purpose ($C$) and bare metal ($BM$) compute cores free for running the application tasks and the operating system (full-fledged Linux). The NIC connects internally with the compute cores through a standard AXI bus; as typical for a NIC, internal packet queues keep a memory pointer to the data structures that actually store the L2 frames, i.e., the header of a Ethernet frame and its payload (protocols from L3 up on). In such a distributed system, we rely on standard network protocols: at L2 Ethernet is used, while on top of it, TCP/IP has been considered to guarantee flow control over packet transmissions. This lead us to the point of choosing an appropriate schema for allocating both MAC addresses (L2) and IP addresses (L3). Indeed, the connection within any pair of compute nodes becomes a dedicated point-to-point network segment. The idea in this case is that of assigning to each compute node in the network a \textit{unique identifier} which is composed of its absolute position in the overall grid (i.e., $\langle R_c, C_c \rangle$; assuming up to $N_c$ nodes in a row and up to $N_r$ rows, the indexes are respectively $R_c \in (0, ..., N_r-1)$ and $C_c \in (0, ..., N_c-1)$). Thus, we map this unique identifier both on the lower bytes of the MAC address string, and on the lower bytes of the IP address. Then, we set the subnet mask in such a way the two lower bytes are used to identify the nodes within the same network. Further, for easing the identification of a compute node in the grid, the unique identifier is reverted to a tuple $ID = \langle G_{rc}, G_{cc}, L_{rc}, L_{cc} \rangle$ at the application level, where $G_{rc}$ and $G_{cc}$ identify the tile within the system, while $L_{rc}$ and $L_{cc}$ identify the compute node within the tile. 
In our system, we exploit the regularity of the topology to implement a dimension order routing (DOR) scheme in software; to this end, we define a kernel module which makes use of netfilter to intercept ingress packets and make decision upon their final destination. In the case the current receiving node is not the final destination, the destination MAC address is changed with that of the neighbourhood interface selected by the DOR routing scheme; then, the frame moves to the transmission data path of the NIC without further intervention of the kernel software stack. Despite its simplicity, DOR routing is effective in distributing the traffic, and can be easily extended to provide more robustness against link faults and traffic congestion. Indeed, the routing scheme may include the capability of switching among alternative source-destination paths, while keeping track of the last traversed link and selecting an alternative output port out of the faulty one should be enough to address links' faults.  

The programmability of the system can be made easy by defining a software runtime that provides primitives to send and receive messages from a specific node, as well as to configure the bandwidth allocation among different transmitting queues:
\begin{itemize}
\item \texttt{send\_msg(*data, size, dst)}: transfers \texttt{size} bytes pointed by \texttt{*data} to a destination node; the destination is selected by encoding the destination $ID$ tuple into a 32~bit integer (\texttt{dst}). If \texttt{size} is greater than the maximum payload for a single frame, the data is automatically split as a sequence of frames.
\item \texttt{recv\_msg(*buf, size, src)}: receives \texttt{size} bytes from a source node; this latter is selected by encoding its $ID$ tuple into a 32~bit integer (\texttt{src}). The content of the received message is stored into a local buffer pointed by \texttt{*buf}, while the call is blocking (it waits until all the \texttt{size} bytes are received or an error condition is raised up).
\item \texttt{set\_conf(*cfg)}: allows to configure the transmission schedule (see Section~\ref{sec:implementation}) of the (local) node; the configuration is pointed by \texttt{*cfg}.
\item \texttt{get\_conf(*buf)}: stores current transmission schedule of the (local) node in the buffer pointed by \texttt{*buf}.
\end{itemize}
On the receiving side, upon receiving a new frame from the network, a hardware module computes the frame checksum to verify if the frame must be dropped (transmission failed) or it has to be moved on in the NIC. Conversely, on the transmission side, the baseline implements a simple round-robin scheduler, which assigns the same fraction of transmission time to each (frame) queue. It is worth to highlight that in our target system, the host system is represented by the complex of general purpose cores $C$. This said, it is still possible to extend the target by hosting distributed nodes within general purpose server machines. In this scenario, the interaction between the local processor (core complex) of the NIC and the main server would be mediated by transactions on a dedicated local bus (e.g., PCIe); however, this scenario is out of the purpose of this work.

% sect. 4
\section{Implementation}
\label{sec:implementation}
The basic NIC hardware that comes with Corundum~\cite{b05} implements a simple round-robin scheduling scheme for the transmission of frames from the multiple available queues (see Section~\ref{sec:backgrounds}). This basic scheduler does not implement any time-aware mechanism to move from one frame queue to the next, thus not providing any mechanism to control the transmission bandwidth of the frame queues. Further, in the basic implementation there is any connection with a prioritization scheme at the application level, thus limiting the users in the way they can control the traffic in the targeted system. 

\begin{figure}[ht]
  \centering
  \includegraphics[width=0.86\linewidth]{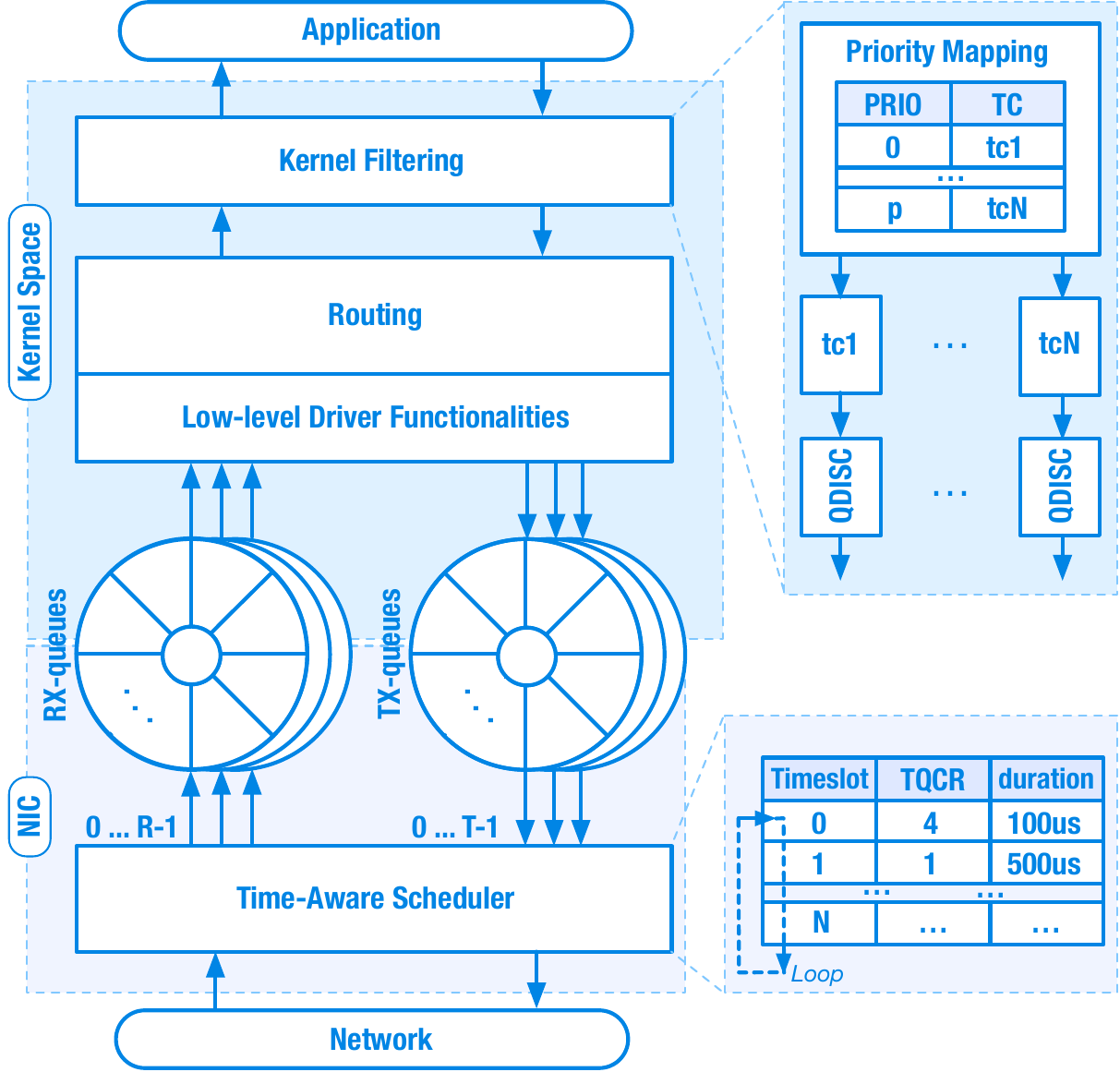}
  \caption{Internal organization of the NIC: time-aware scheduling is implemented in hardware, while priority mapping is enabled at the kernel level, where \textit{prio} is the priority level (0 is the lowest priority) and \textit{TC} are the traffic classes.}
    \label{fig:time_aware_schedule}
\end{figure}

To overcome these limitations, we enabled a time-aware scheduler for the frame transmission and we coupled it with the Linux traffic priority management framework (QDISC) -- see figure~\ref{fig:time_aware_schedule}. 
The implemented solution works independently for each available port on the node. At the basis of our implementation there is the enablement of the hardware management of the precision time protocol (PTP -- IEEE 1588) which allows nodes distributed over a conventional LAN network to keep their internal clocks synchronized. Thus, the internal reference timer (also for the Linux kernel to manage timing events) becomes the one synchronized through the PTP over the point-to-point connections. This internal clock reference is periodically adjusted and guarantees to keep time alignment among neighbourhood nodes with a precision in the order of few tens of nanoseconds. 
The second building block of our implementation is the \textit{time-aware frame queues scheduler}, which is the hardware module of the NIC responsible for selecting the queue from which to transmit, based on a configured time schedule. To this purpose, we enabled the Corundum basic scheduler to loop over the available transmitting queues in such a way, the duration of the queue timeslot (i.e., the time for which the scheduler transmit from a specific queue) is set up from a control register (timeslot queue control register --TQCR). Also, the looping order can be configured through dedicated control registers (schedule control registers --SCRs). The TQCRs implementation provides a microsecond granularity, as well as all the configuration registers are exposed to the software level on the AXI bus. The SCRs and TQCRs form a reconfigurable table through which the software stack can implement specific traffic management policies. An important aspect to keep the frame transmission safer is the addition of a guardband to each timeslot to avoid switching from one queue to another in-between the transmission of a frame. All the frames waiting on the queues not time-scheduled are scheduled for transmission in a round-robin fashion if there is free unused time on the programmed time-aware schedule. Also, if a time-aware frame queue is empty, the scheduler will stuck on that queue for the programmed timeslot, in order to enforce the time-aware priorities.  

The last component for supporting time-aware priorities is enabling the Linux kernel to map traffic priorities onto time-aware scheduled frame queues. Since the configuration registers allow to assign specific timeslots to a specific group of transmitting queues, we selected a subgroup of transmitting queues upon which we associated the SCRs and TQCRs. Then, we used the Linux QDISC tool to make the appropriate mapping. This tool allows the user to map different levels of traffic priority with different transmitting queues passing through potential multiple traffic classes (these are a way to manage the traffic from different queues using a single policy --i.e., the queue discipline (QDISC)). Given this multi-step mapping scheme, we reverted to a 1-to-1 mapping between priorities and traffic classes, and to a 1-to-1 mapping between traffic classes and the time-aware scheduled queues. The lower priority is generally identified with the value 0 and it also maps all traffic which has no specific priority set up.  

\subsection{Preliminary Experimental Evaluation}
\label{subsec:evaluation}
In order to validate our system and without loose of generality with regards to the target system described in Section~\ref{sec:system_overview}, we focused and experimented with a small testbed, composed of a $2 \times 2$ tile connected to other 2 additional nodes on the horizontal dimension. In this way, we replicated a system where we enabled the communication among different tiles, while reducing the overall number of compute nodes.
\begin{figure}[ht]
  \centering
  \includegraphics[width=0.90\linewidth]{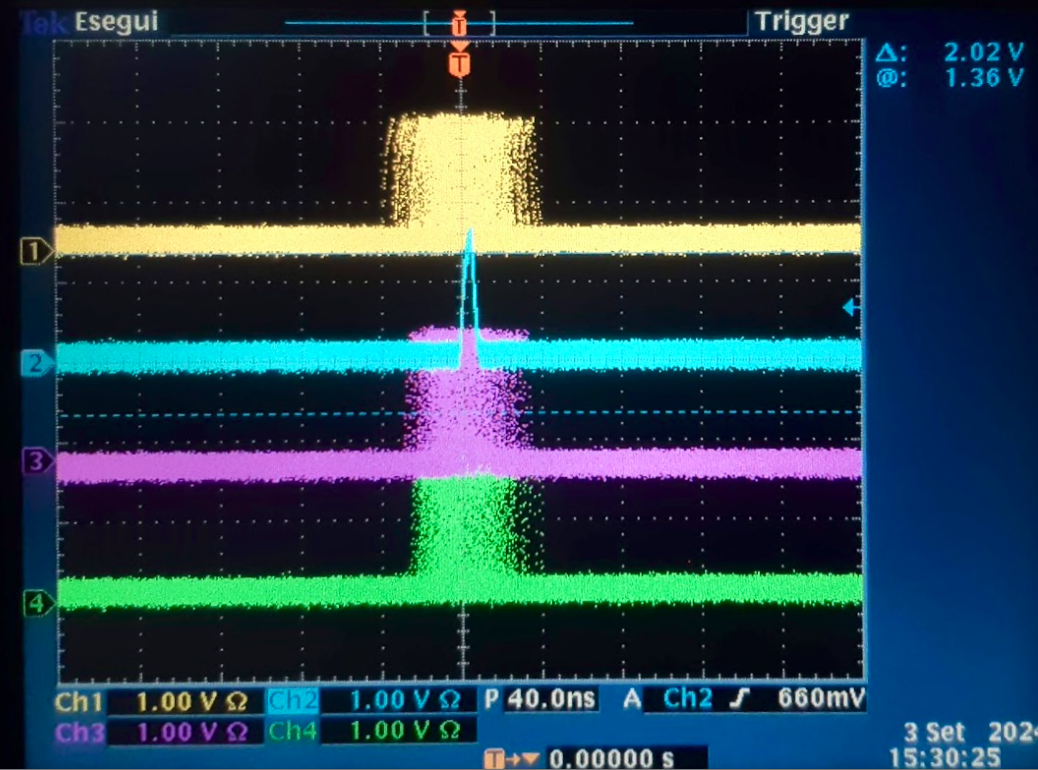}
  \caption{Signal acquisition on 4 SoC FPGA nodes showing the PTP synchronization: the \textit{grandmaster} node (light blue) periodically exchanges PTP-frames with the other \textit{slave} nodes (yellow, green and purple) which adjust their internal clocks to align to the grandmaster. According to the timescale, clocks' variation is in the range of $\pm 40$ w.r.t. the grandmaster.}
    \label{fig:ptp_sync}
\end{figure}
Although this experimental setup is limited in terms of absolute numbers of connected nodes, it is worth to note that it is large enough to demonstrate the full capabilities of the proposed NIC architecture. Indeed, each node implements advanced routing functionalities, and its scaling out capabilities are not limited by the size of the chosen setup.
We selected the AMD Zynq UltraScale$^+$ MPSoC (the specific model was the ZCU-102) as the reference hardware device where to implement our hardware-software solution. As so, each node was equipped with an heterogeneous SoC FPGA embedding 4 general purpose ARM A53 cores and 2 real-time ARM R5F cores. Since each node exposes 4 QSFP ports and an additional RJ45 connector, we decided to use this latter to implement the management interface. 

\begin{figure}[ht]
  \centering
  \includegraphics[width=0.85\linewidth]{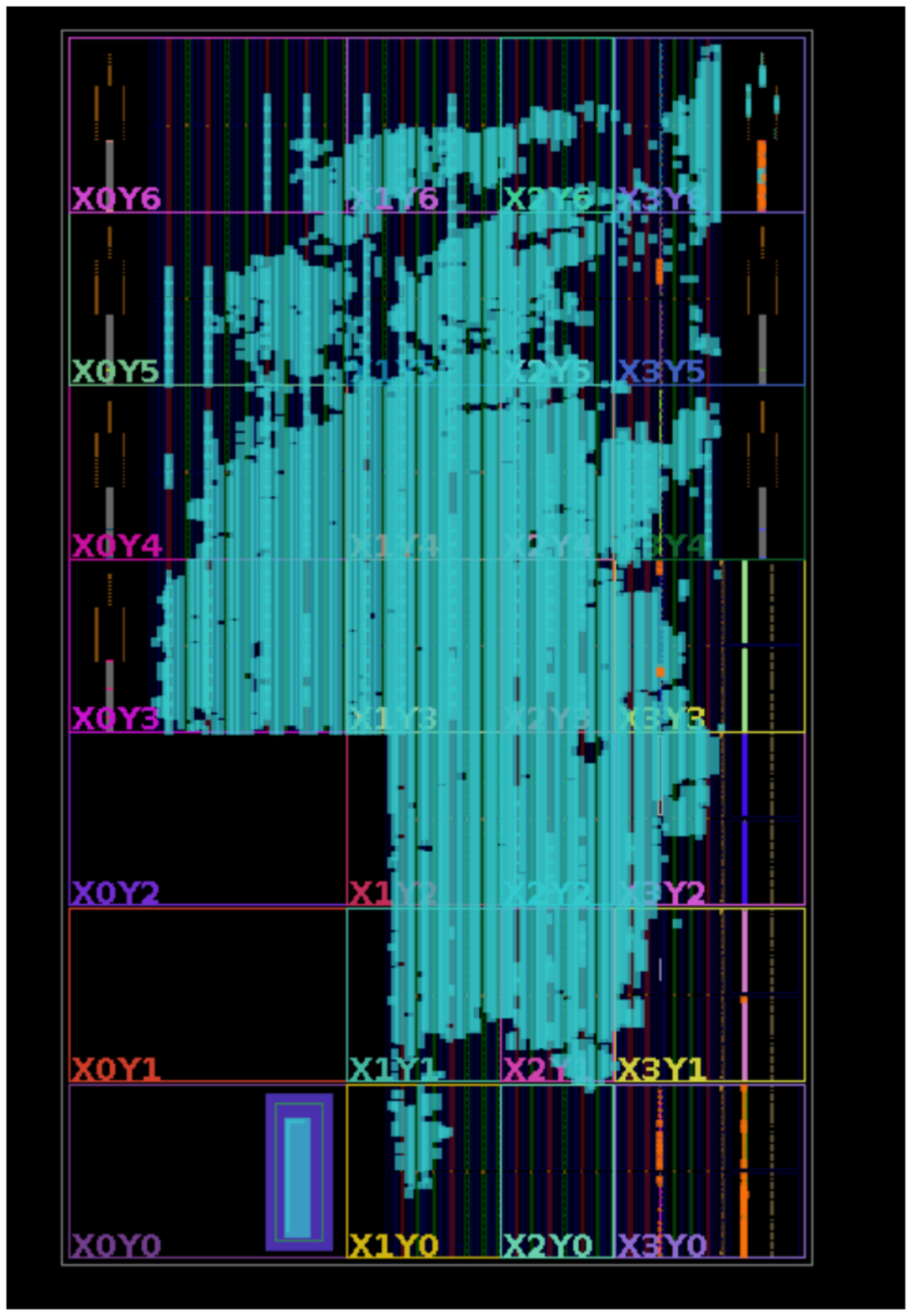}
  \caption{Result of the place and route process after synthesis of the proposed system on the AMD ZCU-102, using AMD Vivado 2023.2.}
    \label{fig:nic_synthesis}
\end{figure}

\begin{table}[ht]
    \centering
    \begin{tabular}{llll}
        \toprule
        \multicolumn{4}{c}{\textbf{System:} AMD ZCU-102}\\
        \midrule
        \textbf{Resource}  & \textbf{Available} & \textbf{Utilization}  & \textbf{Utilization \%} \\
        \midrule
        LUTs    & 274080    & 75042     &  27.38 \\ 
        LUTRAMs & 144000    & 7641      &  5.31  \\
        FFs     & 548160    & 82080     &  14.97 \\
        BRAMs   & 912       & 361.5     &  39.64 \\
        IO      & 328       & 44        &  13.41 \\
        GT      & 24        & 4         &  16.67 \\
        BUFG    & 404       & 20        &  4.95  \\
        MMCM    & 4         & 1         &  25.00 \\
        \bottomrule
    \end{tabular}
    \vspace{0.25cm}
    \caption{Resource utilization after synthesis of the NIC (with the time-aware scheduling of TX-RX queues) on a compute node.}
    \label{tab:resource_synth}
\end{table}
The 4 QSFP ports have been used to connect the nodes each other. We used a server connected to the management switch to automate the startup and configuration of the compute nodes. Each node ran a Linux Ubuntu 20.04 operating system, on top of which the Corundum NIC driver and the routing logic has been implemented, and where we enabled the PTP protocol -- see figure~\ref{fig:ptp_sync}. On the hardware side, we synthesized a new image for the programmable logic, which exports the control registers for the time-aware scheduled transmission queues. Table~\ref{tab:resource_synth} shows the PL resource consumption after the design synthesis for the proposed compute node NIC on the selected device (AMD ZCU-102 SoC FPGA), while figure~\ref{fig:nic_synthesis} shows the results of the place and route process using AMD Vivado 2023.2 design tool. Also, the estimated power consumption is quite low, i.e., $2.840$~W out of $7.532$~W for the programmable logic.
The resulting NIC allowed for transmitting with up to $10$~Gbps on the QSFPs; however, the embedded general purpose cores were not able to saturate this bandwidth, stopping at $2.25$~Gbps (peak bandwidth). We measured this value, experimentally, running a Linux \textit{iperf3} instance on one of the two single nodes connected to the central tile (iperf3 client), and targeting the other single node (iperf3 server). As such, we were able to measure also the communication latency, which averaged $0.35$~ms on the shortest path, and reached $0.52$~ms when simulating a fault on a link on the shortest path. The average value for the jitter was $\sim$$8\mu s$ with a standard deviation of few $\mu s$.
\begin{figure}[ht]
  \centering
  \includegraphics[width=0.90\linewidth]{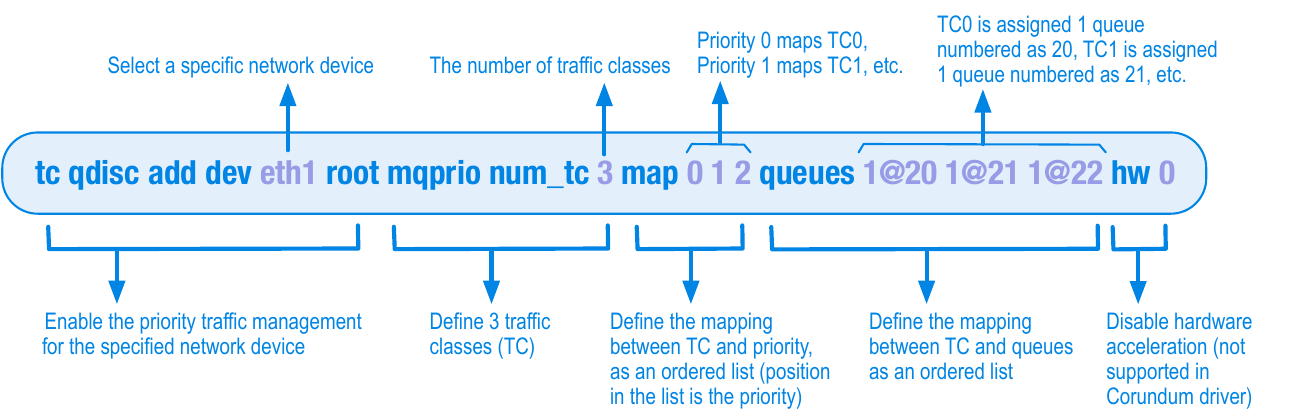}
  \caption{Example of the QDISC command to create the mapping between 3 priorities, 3 traffic classes (TC) and 3 time-aware scheduled queues, on a specified interface (eth1).}
    \label{fig:qdisc_mapping}
\end{figure}

To preliminary test our proposed solution, we mapped 3 priority levels for the traffic ($prio \in \{0, 1, 2\}$) to 3 traffic classes, and we programmed QDISC to map these traffic classes to 3 time-aware scheduled queues (see figure~\ref{fig:qdisc_mapping}). Then, we enforced the queue associated to the highest priority to keep the $90\%$ of the transmission time, while leaving the remaining $10\%$ for the other queues (comprising the other two time-aware queues). Since the selected high priority queue was set to transmit for the $90\%$ of the time, equivalently it saw the $90\%$ of the transmission bandwidth. We verified this by running a Linux \textit{iperf3} instance on one of the two single nodes connected to the central tile and assigning the highest priority, while running another \textit{iperf3} instance with the lowest priority assigned. The evaluation showed that approximately $89\%$ of the peak bandwidth ($\sim$2.00~Gbps) was used by the iperf3 instance mapped on the highest priority queue, demonstrating that we can implement a time-aware priority control.

% sect. 5
\section{State of the Art}
\label{sec:sota}
One of the key elements determining the performance of a distributed computer system is the network. Apart from well-known performance metrics (e.g., bandwidth, latency, jitter), modern applications require a stringent deterministic behaviour, i.e., deterministic transmission of the frames. Time Sensitive Network (TSN) emerged as a set of standards defined upon the IEEE 802.1 to reach low-latency, deterministic traffic management on Ethernet-based networks (IEEE 802.3) for time-sensitive applications like audio/video streaming and real-time control. The majority of the work done by the TSN working groups regards the extension of IEEE 802.1Q standard. Among the TSN projects, scheduling and traffic shaping are of primary importance to provide mechanisms guaranteeing a deterministic behaviour of the network nodes. In particular, the traffic shaping refers to the process of transmitting the frames in such a way the egress traffic is smoothed according to a defined policy. Many of the mechanisms defined by TSN working groups for scheduling and shaping the traffic can be efficiently implemented in hardware.

Albeit many industrial switches and network cards offer TSN-aware functionalities, there are less research works targeting the design, implementation and analysis of TSN-based network equipments, and aimed at providing an open tool for testing new ideas and architectures. Zhou, Berger and Yan~\cite{b16} proposed a FPGA-based implementation of the Asynchronous Traffic Shaper (ATS). The authors also compared their implementation with a more standardized solution based on the strict-priority shaper (SPS). The ATS has been proposed to guarantee a bounded transmitting delay without the need for a global (among the node of the networks) time-aware scheduling of the transmission windows. As such, it provides a local, asynchronous mechanism to dynamically adjust the transmission time for each priority queue (referred to as Interval Priority Value --IPV). The authors describe a FPGA implementation, where ATS is placed on the transmission path (egress traffic). Here, a dedicated task classifies the upcoming traffic stream (frame by frame) and inserts it into dedicated priority queues. Priority levels as defined in the QoS field of the frames are preserved, while the classification is based on a set of rules defined during the design phase. These rules consider the frame header information as well as some other transmission statistics to adjust the transmission \textit{eligibility time}, which defines when a frame from a given priority queue can be transmitted. The ATS iterates over the priority queues and select the frame with the highest IPV that reached its eligibility time. Compared to the simpler SPS (which just iterates over and stuck on a priority queue till there are frames to transmit), the ATS performs bounded frame transmissions at the cost of a much larger FPGA resource consumption and a slower clock frequency. Further, the proposed solution comes without any description of how a Linux environment can be integrated with. 

Similarly, Ben Ahmed et al.~\cite{b19} described a FPGA-based implementation of a TSN-based switching architecture, specifically supporting ATS. Experiments targeted an AMD Kintex 7 FPGA (KC705 Evaluation Kit), which exposes fewer reconfigurable resources with respect to the AMD ZCU-102 devices used in our experiments; the resource occupancy is of $76.46\%$ LUTs, $32.78\%$ FFs and $92.81\%$ BRAMs, which is quite larger than our implementation.

Pruski and Berger~\cite{b18} described a high-performance TSN-aware switching architecture; specifically the authors implemented both the Time Aware Shaper (TAS), which is analogous to our time-aware scheduler, and the Credit Based Shaper (CBS). The most limiting aspect of this work is in the absence of a concrete implementation on a FPGA and its integration with the Linux environment. In~\cite{b17} the authors performed experiments aimed at evaluating the guarantees of latency and jitter enforced by the IEEE 802.1 Qbv standard. Compared to our experimental setup, the authors relied on off-the-shelf components (TSN-compliant industrial switches) that have been programmed to support the time constraints of a (real-time) control application (a video streaming application competing for bandwidth was also executed). This experimental setup lacked of the integration with a Linux environment.

Other works analysed different aspects of TSN-based network systems (primarily, switching architectures) both using software simulation tools~\cite{b20, b21, b22} and hardware implementations~\cite{b23, b24, b26}.

It is worth to highlight that all the aforementioned implementations are bounded to a maximum bandwidth of 1~Gbps (and in some cases to 100~Mbps), which is quite lower than our proposed experimental setup, and that no one targeted the integration with the Linux environment.

%% From previous section -- Backgrounds
Corundum~\cite{b05} has been proposed as an open platform supporting the development and test of new networking protocols and architectures, while performing at realistic line rates. Its modular design allows users to experiments with custom functionalities both at the software and hardware levels. As such, Corundum represents the primary choice for researchers. In~\cite{b06}, the authors relied on the Corundum modular design to extend it through the integration of a custom RISC-V based core complex. They demonstrated that it was possible to easily customize the original design by introducing additional hardware components without impacting on the NIC performance. Similarly, Lin et al.~\cite{b07} exploited the versatility of the Corundum design to implement an effective Match-Action Table (MAT) architecture, supporting up to $8 \times 25$~Gbps line rates. Although the MAT architecture is designed to perform different tasks on in-transit packets, its routing capability is limited, and poorly matches a distributed system where compute nodes are connected through separated point-to-point network segments. On the other hand, the design in~\cite{b06} focuses on making the NIC capable of offloading host CPU on complex data tasks. Other than Corundum, few other projects propose an open hardware-software architecture. In~\cite{b08} the authors proposed superNIC (sNIC), a FPGA-oriented design focused on easing programmability. Cabal et al.~\cite{b25} proposed the Network Development Kit (NDK), a solution similar to Corundum and capable of operating at line rates up to 400~Gbps (on specific hardware configurations).

Besides open research projects, Smart-NICs are quite common in the vendors' catalogues and within hyperscalers' infrastructures~\cite{b10, b11, b12, b13, b14}. Both Nvidia, AMD and Intel have their own models. Nvidia~\cite{b09} and AMD products provides up to 400~Gbps line rates and come with a large set of general purpose cores (running full-fledged Linux OSes) and specialized acceleration engines.

% sect. 6
\section{Conclusion}
\label{sec:conclusion}
In this work, we explored the versatility of the Corundum open platform to enable a priority traffic management with a precise, time-aware scheduling mechanism. The proposed solution exports dedicated control registers for easily programming the traffic scheduler, while the Linux QDISC tool is used to map OS priorities to time-aware scheduled queues. Preliminary tests demonstrated the effectiveness of this approach. 
%Future activities will be focused on acquiring more statistics over a even larger experimental setup, as well as to implement routing functionalities directly in hardware. 

Future activities will be focused on acquiring more statistics over an even larger experimental setup (i.e., including a larger number of tiles). Moreover, we want to consider real applications as traffic sources, as well as different traffic patterns. Another direction of investigation is given by introducing and evaluating optimizations of the proposed design. Specifically, we are targeting, as a future activity, the implementation of a credit-based queue scheduling mechanism and the use of bare-metal cores to execute routing functionalities more efficiently and fully exploiting link bandwidth. Last but not least, as a future research activity, we also want to compare our implementation with State-of-the-Art alternatives.

% references

% ------------------ END OF DOCUMENT -------------------
\end{document}